\documentclass[12pt]{article}
\usepackage{amsmath,amsfonts,amssymb,a4,epsfig}

\newcommand{\R}{{\mathbb{R}}}

\def\im{{\rm Im}}

\def\ra{\rightarrow}
\def\preuve{\begin{proof}}

\def\gg{\gamma}

\def\gl{\lambda}
\def\go{\omega}

\def\gs{\sigma}

\newtheorem{theo}{Theorem}

\begin{document}

\title{Scattering and correlations}
\author{Yves Colin de Verdi\`ere \footnote{Institut Fourier,
 Unit{\'e} mixte
 de recherche CNRS-UJF 5582,
 BP 74, 38402-Saint Martin d'H\`eres Cedex (France);
yves.colin-de-verdiere@ujf-grenoble.fr
}}


\maketitle



\section*{Introduction}
Let us consider the propagation of scalar waves with the speed $v>0$
given by the 
wave equation $u_{tt}-v^2 \Delta u =0$
outside a compact domain $D$ in
the Euclidean space $\R^d $.
Let us put $\Omega = \R^d \setminus D $.
 We can assume for example
Neumann boundary conditions. We will denote by $\Delta _\Omega $
the (self-adjoint) Laplace operator with the boundary conditions.
So our stationary wave equation  is the Helmholtz equation
\begin{equation} \label{equ:helmholtz}
 v^2 \Delta _\Omega f + \go^2 f=0  
\end{equation}
 with the boundary conditions.
We consider a bounded interval $I=[\go _-^2, \go _+^2 ]\subset ]0,+\infty [$
and the Hilbert subspace ${\cal H}_I $ of $L^2(\Omega )$  which is the image 
of the spectral projector $P_I $ of our  operator  $-v^2 \Delta _\Omega $.

Let us compute the integral kernel  $\Pi_I (x,y)$  of $P_I$
defined by:
\[ P_I f(x)=\int _\Omega \Pi _I (x,y) f(y) |dy| \] 
 into 2 different ways:
\begin{enumerate}
\item From general spectral theory
\item From  scattering theory.
\end{enumerate}
Taking the derivatives of $\Pi_I (x,y)$  w.r. to $\go _+$, we get
a simple general and exact relation
between the correlation of scattered waves and the Green's function
confirming the calculations from \cite{SS}
in the case where $D$ is a disk. 
See also \cite{YCdV2} for other relations between correlations and
Green's functions used in passive imaging. 

\section{$\Pi_I(x,y)$             from spectral theory}

Using the resolvent kernel (Green's function)
$G(\go ,x,y)=[ (\go ^2  + v^2\Delta_\Omega ) ^{-1 } ](x,y) $
for $\im \go  >0 $
and the Stone formula, we have:
\[ \Pi_I(x,y)=-\frac{2}{\pi }\im \left(\int _{\go_-}^{\go_+}
G(\go+i0, x,y) \go d\go \right)  \] 
Taking the derivative w.r. to $\go_+ $ of $\Pi_{[\go _{-}^2 ,\go_{+}^2]}(x,y)$,
we get 
\begin{equation} \label{equ:stone}
 \frac{d}{d\go} \Pi_{[\go _-,\go^2]} (x,y)=-\frac{2 \go}{\pi }\im (
G(\go+i0, x,y))  ~. \end{equation}

\section{Short review of scattering theory}

They are many references for scattering theory: for example
\cite{Ra,R-S}.

Let us define for ${\bf k} \in \R^d$  the plane wave
\[ e_0 (x,{\bf k})=e^{i<{\bf k}|x> }~.\]
We are looking for solutions  
\[ e (x,{\bf k})= e_0 (x,{\bf k})+ e^s(x,{\bf k})\]
of the 
Helmholtz equation (\ref{equ:helmholtz})
 in $\Omega $
 where $e^s$, the scattered wave satisfies the so-called
Sommerfeld radiation condition\footnote{As often, we denote
$k:=|{\bf k}|$ and $\hat{{\bf k}}:={{\bf k}}/{k}$}:
\[  e^s(x,{\bf k})=\frac{e^{ik|x|}}{|x|^{(d-1)/2}}\left( e^{\infty}
(\frac{x}{|x|},{\bf k }) + O(\frac{1}{|x|}) \right),
~x\ra \infty ~.\]
The complex function $e^{\infty}
(\hat{x},{\bf k })$ is usually called the 
{\it scattering amplitude}.

It is known  that the previous problem admits an unique solution.
In more physical terms, $e(x,{\bf k})$ is the wave generated by the full
scattering process from the  plane wave $e_0 (x,{\bf k})$. 
Moreover we have a generalized Fourier transform:
\[ f(x)=(2\pi )^{-d}\int_{\R ^d} \hat{f}({\bf k }) e (x,{\bf k})|d{\bf k}|\]
with 
\[ \hat{f}({\bf k })=\int_{\R ^d}  \overline{e (y,{\bf k})}f(y) |dy|~. \]

From the previous generalized Fourier transform, we can get the
kernel of any function $\Phi (-v^2 \Delta _\Omega )$ as follows:
\begin{equation} \label{equ:function}
 [\Phi (-v^2 \Delta _\Omega )](x,y)=
(2\pi )^{-d}\int_{\R ^d}\Phi (v^2 k^2) e (x,{\bf k})\overline{e (y,{\bf k})}
|d{\bf k}|~.\end{equation}

\section{$\Pi_I(x,y)$    from scattering theory}

Using  Equation (\ref{equ:function}) with
$\Phi =1_I $ the 
 characteristic functions
of some bounded interval $I=[\go_-^2,\go^2]$,
  we get:
\[ \Pi_I (x,y)=
(2\pi)^{-d}\int _{\go _- \leq vk  \leq \go  } e(x,{\bf k})
\overline{e(y,{\bf k})} |d {\bf k}| ~.\]
Using polar coordinates and defining $| d\gs |$
as the usual measure on the unit $(d-1)-$dimensional sphere, we get:
\[ \Pi_I (x,y)=(2\pi)^{-d}\int _{\go _- \leq v k \leq \go  }k^{d-1}dk
\int _{{\bf k}^2=k^2 }
 e(x,{\bf k})
\overline{e(y,{\bf k})} | d\gs | ~.\]
We will denote by $\gs _{d-1}$ the total  volume
 of the unit sphere in $\R^d$: $\gs _0=2,~\gs _1=2\pi,~\gs _2= 4\pi,\cdots $.

Taking the same derivative as before, we get:
\[  \frac{d}{d\go } \Pi_{[\go _-^2,\go ^2]}
  (x,y)=(2\pi)^{-d}\frac{\go^{d-1}}{v^d}\int _{vk=\go }
 e(x,{\bf k})
\overline{e(y,{\bf k})} |d\gs| ~.\] 

Let us look at $e(x,{\bf k})$ as a random wave with
$k=\go/v$ fixed.
The point-point correlation of such a random wave
 $C_\go ^{\rm scatt}(x,y)$
is given by:
\[  C_\go ^{\rm scatt}(x,y)=\frac{1}{\gs _{d-1}}\int _{vk=\go }
 e(x,{\bf k})
\overline{e(y,{\bf k})} |d\gs |. \]
Then we have:
\begin{equation}  \label{equ:eigen}
 \frac{d}{d\go } \Pi_{[\go _-^2,\go ^2]} (x,y)=(2\pi)^{-d}
\frac{\go^{d-1}\gs_{d-1}}{v^d}
  C_\go ^{\rm scatt}(x,y)~. \end{equation}

\section{Correlation of scattered plane waves and Green's function:
the scalar case}

From Equations (\ref{equ:stone}) and (\ref{equ:eigen}), we get:
\[(2\pi)^{-d} \frac{\go^{d-1}\gs _{d-1}}{v^d} C_\go ^{\rm scatt}(x,y)=
                   - \frac{2\go}{\pi }\im (
G(\go +i0, x,y))  ~. \]
Hence, we have 
\begin{theo} For the scalar wave equation $u_{tt}-v^2 \Delta u =0$
outside a bounded domain in $\R^d$, we
  have the following expression of the correlation of scattered wave
  of
frequency $\omega $ in terms of the Green's function:
\[  C_\go ^{\rm scatt}(x,y)=-\frac{2^{d+1}  \pi ^{d-1}
 v^d }{\gs _{d-1} \go ^{d-2} }
\im (G(\go +i0,x,y))~. \]
\end{theo}
For later use, we put
\begin{equation}
\label{equ:gamma}
 \gg_d=\frac{2^{d+1} \pi ^{d-1}  }{ \gs _{d-1}}~.
\end{equation}

\section{The case of elastic waves}

We will consider the  elastic wave equation
in the domain $\Omega $:
\[ \hat{H}{\bf u} - \go ^2  {\bf u} =0, \]
with self-adjoint  boundary conditions.
We will assume that, {\it at large distances,} we have
\[ \hat{H}{\bf u}=-a ~\Delta {\bf u} - b ~{\rm grad}~{\rm div }{\bf u}~. \]
where $a$ and $b$ are constants:
\[ a= \frac{\mu}{\rho},~b=\frac{\gl + \mu}{\rho}\]
with $\gl,~\mu $ the Lam\'e's coefficients and $\rho $
the density of the medium. We will denote 
$ v_P:=\sqrt{a+b}$ (resp. $v_S:=\sqrt{a}$)
the speeds of the $P-$(resp. $S-$)waves near infinity.

\subsection{The case $\Omega = \R ^d $}

We want to  derive the spectral decomposition of $\hat{H}$ from
the Fourier inversion formula. Let us choose,  for ${\bf k}\ne 0 $,
by $\hat{\bf k}, \hat{\bf k}_1, \cdots, \hat{\bf k}_{d-1}  $
 an orthonormal basis of  $\R ^d $ with $\hat{\bf k}=\frac{\bf k}{k}$
such that these vectors depends in a measurable way of ${\bf k}$.
Let us introduce 
$P_P^{\bf k}=\hat{\bf k}\hat{\bf k}^\star $ the orthogonal projector
onto $\hat{\bf k}$ and $P_S^{\bf k}
=\sum_{j=1}^{d-1} \hat{\bf k}_j\hat{\bf k}_j^\star $
so that $P_P + P_S= {\rm Id}$.
Those projectors correspond respectively to the polarizations
of $P-$ and $S-$waves. 

We have 
\[\begin{array}{l}
 \Pi _I (x,y)=
(2\pi)^{-d}
\int _{\go ^2 \in I} \go ^{d-1} d\go \left( 
v_P^{-d}\int _{v_P k =\go}e^{i{\bf k}(x-y)}P_P^{\bf k} d\gs
+\right. \\
\left. v_S^{-d}\int _{v_S k =\go }e^{i{\bf k}(x-y)}
P_S^{\bf k} d\gs \right) ~. \end{array} \]       
using the plane waves
\[ e_P^O (x,{\bf k})=e^{i{\bf k}x} \hat{\bf k} \]
and 
\[ e_{S,j}^O (x,{\bf k})=e^{i{\bf k}x} \hat{\bf k}_j \]
we get the formula\footnote{We use the ``bra-ket'' notation 
of quantum mechanics:  $|e\rangle \langle f| $
is the operator $x\ra \langle f|x \rangle e $
where the brackets are linear w.r. to the second entry and anti-linear
w.r. to the first one}:
\[ \begin{array}{l} \Pi _I (x,y)=
(2\pi)^{-d}
\int _{\go ^2 \in I} \go ^{d-1} d\go \left( 
v_P^{-d}\int _{v_P k=\go}  |e_P^O (x,{\bf k})\rangle \langle 
 e_P^O (y,{\bf k})|  d\gs
+\right.\\\left.
v_S^{-d}\sum_{j=1}^{d-1} \int _{v_S k=\go }
  |e_{S,j}^O (x,{\bf k})\rangle \langle  e_{S,j}^O (y,{\bf k})|
 d\gs \right) ~. \end{array}       \]

\subsection{Scattered plane waves}

There exists scattered plane waves
\[  e_P (x,{\bf k})= e_P^O (x,{\bf k})+ e_P^s (x,{\bf k})\]
\[ e_{S,j} (x,{\bf k})= e_{S,j}^O (x,{\bf k})+ e_{S,j}^s (x,{\bf k})\]
satisfying the Sommerfeld condition and from which we can deduce
the spectral decomposition of $\hat{H}$.

\subsection{Correlations of scattered plane
 waves and Green's function}

Following the same path as for scalar waves, we get an 
 identity which holds now for the full Green's tensor
$\im {\bf G} (\go +iO, x,y)$:
\begin{theo} For the elastic wave equation, we have the following
  expression of the imaginary part of the Green's function in terms
of the correlation of the scattered S- and P- waves:
\[\begin{array}{l} \im {\bf G} (\go +iO, x,y)=-\gg_d^{-1} \go^{d-2} \left(
\frac{1}{\gs _{d-1} v_P^{d}}\int _{v_P k=\go }  |e_P (x,{\bf k})\rangle 
 \langle e_P (y,{\bf k})|  d\gs\right.
+\\\left. 
\frac{1}{\gs _{d-1} v_S^{d}}\sum_{j=1}^{d-1} \int _{v_Sk=\go}
  |e_{S,j} (x,{\bf k})\rangle 
 \langle e_{S,j} (y,{\bf k})|  
 d\gs \right)
 ~,\end{array} \]
with $\gg_d $ defined by Equation (\ref{equ:gamma}).
\end{theo}

This formula expresses the fact that the correlation of scattered 
plane waves randomized with the appropriate weights ($v_P^{-d}$
versus $v_S^{-d}$) is proportional to the
Green's tensor.  Let us insist on the fact that this true everywhere 
in $\Omega $ even in the domain where $a$ and $b$ are not constants.

\bibliographystyle{plain}

\end{document}